\documentclass[twocolumn,showpacs,amsmath,amssymb,prd,nofootinbib]{revtex4}
\usepackage{epsfig}
\def\beq{\begin{equation}}
\def\eeqno#1{\label{#1}\end{equation}}

\def\rarrow{\rightarrow }

\def\dleft{\rlap{{\it D}}\raise 8pt
\hbox{$\scriptscriptstyle\Leftarrow$}}
\def\dright{\rlap{{\it
D}}\raise 8pt\hbox{$\scriptscriptstyle\Rightarrow$}}

\def\az{a_{0}}

\def\l0{\ell_{0}}

\def\a{\alpha}
\def\b{\beta}
\def\l{\lambda}

\def\k{\kappa}

\def\n{\nu}

\def\vinf{V\_{\infty}}

\def\a{\alpha}
\def\xlimin{{x\rarrow\infty \atop{\raise 1pt\hbox to 30pt
{\rightarrowfill}}}}
\def\limlim#1#2{{#1\rarrow #2 \atop{\raise 1pt\hbox to 30pt
{\rightarrowfill}}}}

\def\vv{{\bf v}}

\def\S{\Sigma}

\def\Q{\mathcal{Q}}

\def\gN{g\_N}

\def\a{\alpha}
\def\b{\beta}

\def\n{\nu}

\def\_#1{_{\scriptscriptstyle #1}}
\def\^#1{^{\scriptscriptstyle #1}}

\def\vsav{\langle V^2\rangle}
\begin{document}
\title{Global Deep-MOND Parameter as a Theory Discriminant}
\author{Mordehai Milgrom}
\affiliation{Department of Particle Physics and Astrophysics, Weizmann Institute}

\begin{abstract}
Different formulations of MOND predict somewhat different rotation curves for the same mass distribution. Here I consider a global attribute of the rotation curve that might provide a convenient discriminant between theories when applied to isolated, pure-disk galaxies that are everywhere deep in the MOND regime. This parameter is $\Q\equiv\vsav/\vinf^2$, where $\vsav\equiv M^{-1}\int 2\pi r\S(r)V^2(r)dr$, with $\S(r)$ the disk's surface density, $M$ its total mass, and $\vinf$ is the asymptotic (constant) rotational speed.
The comparison between the observed and predicted values of $\Q$ is oblivious to the distance, the inclination, the mass, and the size of the disk, and to the form of the interpolating function. For the known modified-gravity theories $\Q$ is predicted to be a universal constant [independent of $\S(r)$]: $\Q=2/3$. The predicted $\Q$ value for modified-inertia theories does depend on the form of $\S$. However, surprisingly, I find here that it varies only little among a very wide range of mass distributions, $\Q\approx 0.73\pm 0.01$. While the difference between the theories amounts to only about 5\% in the predicted rms velocity, a good enough sample of galaxies may provide the first discerning test between the two classes of theories.
\end{abstract}
\pacs{04.50.-h  98.52.Eh  98.80.-k}
\maketitle
\section{Introduction}
In applying modified Newtonian dynamics (MOND), we use several formulations and theories (for a recent review of MOND and its formulations see \cite{fm12}). The nonrelativistic formulations in use are the nonlinear version of the Poisson equation \cite{bm84}, quasilinear MOND (QUMOND) \cite{milgrom10a}--both of which may be classified as modified-gravity (MG) theories--and a class of so-called `modified-inertia' (MI) theories \cite{milgrom94a,milgrom11}.
\par
We do not know yet which of these formulations, if any, is in the right direction. It is important to try and decide between these options, because they each point to a different direction in constructing relativistic formulations, and, more generally, in pinpointing the deeper origins of MOND. Of the known relativistic formulations of MOND, TeVeS \cite{bek04}, MOND adaptations of Einstein-Aether theories \cite{zlosnik07}, most versions of bimetric MOND (BIMOND) \cite{milgrom09}, and those based on a polarizable medium \cite{blt09} have a version of the nonlinear Poisson theory as their nonrelativistic (NR) limit; while certain versions of BIMOND tend to QUMOND in the NR limit.
\par
But, to my knowledge, no systematic endeavor has been undertaken to observationally discriminate between the different theories. The main reason for this must be that despite the
pronounced conceptual differences between them, the differences between their salient predictions are small: by and large, these predictions follow directly from the basic tenets of MOND, which all formulations share. These tenets are \cite{milgrom09a}: departure from Newtonian dynamics for accelerations  $a\lesssim\az$, and space-time scale invariance in the deep-MOND limit (DML) $a\ll \az$, where $\az$ is the MOND acceleration constant.
\par
Arguably, the most effective way to distinguish observationally between the theories would be based on their somewhat different predictions of rotation curves (RCs) of disk galaxies. This, however, is not so easy. For all galaxies, the basic tenets alone dictate that the rotational speed becomes constant at large radii, and they also fix the value of this speed. For galaxies with high accelerations near their centers, all theories predict Newtonian speeds there, and the basic tenets predict that the transition from Newtonian to MOND behavior should occur around the radius where $V^2(r)/r=\az$. So the predicted curves follow almost in full from only the basic tenets. For galaxies with $a\ll\az$ everywhere, we shall see below that the differences between theories of the  predicted rms values of the rotational speed are only about 5 percent. Indeed, Ref. \cite{brada95} showed through a few numerical examples that the predicted differences between the nonlinear-Poisson formulation of MG, and MI theories, are of this order.
\par
In addition, some of the differences between the predictions may be reduced by choosing different forms of the interpolating function that appears in all theories, and also by
assuming different stellar mass-to-light (M/L) values (which are not known accurately) when predicting the RC of a given galaxy.
 \par
Because of these hindrances, one would require a large sample of galaxies with very good quality data to conduct meaningful tests. But the task of computing RCs for many galaxies in MG theories is quite onerous. Perhaps because of all the above obstacles, no attempt has been made so far to conduct a systematic discriminating study.
All but a few MOND RC analyses of real galaxies to date were performed using the rather more manageable MI predictions. An exception is the recent analysis of several RCs using QUMOND \cite{angus12}.
\par
Here, I discuss a useful shortcut for performing such an analysis using not the full RC, but a global parameter predicted to attain different values by different theories. Its main advantage is that it eliminated the need to compute RCs for different galaxies, since, as I show here, it takes an almost universal value within each theory class.

\section{The $\Q$ parameter}
It follows from the basic tenets of MOND, and hence in any present MOND theory, that the asymptotic circular rotation speed around any isolated mass is constant and depends only on the total mass $M$:
\beq \vinf=(MG\az)^{1/4}. \eeqno{plama}
(Strictly speaking, only a proportionality is dictated, but $\az$ is normalized so as to give equality.)
Consider a pure-disk galaxy of surface density $\S(r)$ and RC $V(r)$, that is wholly in the deep-MOND limit (DML): $V(r)^2/r\ll\az$ everywhere. We define
 \beq \Q\equiv \frac{\vsav}{\vinf^2}=\frac{1}{M\vinf^2}
 \int_0^\infty 2\pi r\S(r)V^2(r)dr \eeqno{mipa}
($\vsav$ here is not to be confused with the velocity dispersion).
When testing MOND itself it behooves us to work with the velocities themselves since the amplitude of the RC is also a prediction of MOND. But since all MOND theories predict the same asymptotic speed, we gain much, when only comparing MOND theories, by considering the normalized speeds as in the definition of $\Q$.
I restrict myself to pure disks because otherwise the unknown kinematics of the bulge component would enter. Restriction to the DML has the following advantages: (i) It frees us from dependence on the exact form of the interpolating function, since the DML form of this function is the same for all theories. (ii) Because of the scaling properties of the DML, and the choice of normalized speeds in $\Q$, its predicted values depend only on the form of $\S(r)$, not on the mass and size of the galaxy, predicting the same value for all surface densities $\k\S(r/h)$ independent of $\k$ and $h$. This means that comparison of the predictions with the data is oblivious to the assumed distance to the galaxy or its (overall) inclination. This is not the case for disks that are not in the DML.
(iii) In this limit the known MG formulations of MOND predict a universal value of $\Q$, saving us the need for onerous numerical solutions of the nonlinear potential problem for each galaxy separately. (iv) A surprising result of the present work is that the $\Q$ value predicted for MI theories is also almost universal (while different from that predicted by MG theories).

\subsection{The predicted $\Q$ value--modified-gravity theories}
The two nonrelativistic, MG theories in use today, the nonlinear Poisson formulation and QUMOND probably predict somewhat different RCs for the same mass distribution (although I am not aware of any actual comparison between the two). However, they predict the same DML virial relation:
for any isolated, stationary, self-gravitating DML system made of masses $m_i$,  both theories predict \cite{milgrom94b,milgrom10a}
 \beq \vsav=\frac{2}{3}(MG\az)^{1/2}(1-\sum q_i^{3/2}), \eeqno{mapapa}
where $\vsav\equiv M^{-1}\sum m_i\vv_i^2$ is the mass weighted mean-squared (three-dimensional) velocity in the system (the velocities $\vv_i$ are measured with respect to the center of mass), and
$q_i=m_i/M$ are ratios of particle masses to the total mass.\footnote{In calculating $\vsav$ the internal motions within the masses $m_i$ are immaterial. If we want to include those we need to consider the constituents of $m_i$ as elementary masses, and ensure that their internal dynamics are in the DML. If they are not (as in stars), or if $m_i$ themselves are not even self-gravitating (as in atoms), we need to consider them as elementary bodies.}$^,$\footnote{It does not matter if the accelerations inside or in the near vicinity of $m_i$ are high. It is only required that the `mean-field' accelerations are small everywhere.}
Applying this to a disk galaxy, where $m_i$ are stars or gas clouds, we have $\sum q_i^{3/2}\ll 1$: if the system is made of $N$ masses with $q_i\sim 1/N$ then $\sum q_i^{3/2}\sim N^{-1/2}$. We thus have from Eqs. (\ref{plama})-(\ref{mapapa}), in the large-$N$ limit,
 \beq \Q=2/3,  \eeqno{mushya}
 with a correction of order $N^{-1/2}$.
\par
So both theories predict the same rms velocity (and the same asymptotic speed) for DML galaxies, which must mean that their predicted RCs are very similar.

\subsection{The predicted $\Q$ value--modified-inertia theories}
We do not yet have an example of a full-fledged MI theory \cite{milgrom94a,milgrom11}, but, interestingly, we do know exactly the form of the predicted RC for all such theories.
A general theorem \cite{milgrom94a} states that for circular orbits in axisymmetric potentials (applicable to the circular motions in disk galaxies) the MOND acceleration, $g(r)$, and the Newtonian one, $\gN(r)$, at radius $r$, are related algebraically: $g(r)=\gN(r) \n[\gN(r)/\az]$, where $\n(y)$ is a function that depends only on the theory (it is derived from the action of the theory restricted to circular orbits). The basic tenets of MOND dictate the asymptotic behavior of $\n(y)$ as small and large arguments. In particular, they dictate the unique DML relation universal to all MI theories:
 \beq g(r)=[\az g\_N(r)]^{1/2}.  \eeqno{palupa}
(For some disks $\gN$ may point outward at some $r$; then $g$ does as well; but I ignore this for simplicity.)
So, the predicted value of $\Q$ can be written in terms of the Newtonian RC of the disk, $V_N(r)$:
 \beq \Q=\frac{2\pi}{M(MG)^{1/2}}\int_0^{\infty} r^{3/2}\S(r)V_N(r)dr. \eeqno{litaya}
This does not lead to a general virial relation of the type of Eq.(\ref{mushya}): the value of $\Q$ is not universal.
\par
To see if $\Q$ can serve as a useful discriminant, I calculated its value for various classes of $\S(r)$ to see how similar they are to each other, and how different they are from the universal $\Q=2/3$ of MG theories.
\par
For Kuzmin disks having $\S(r)\propto [1+(r/h)^2]^{-3/2}$
 I find
 \beq \Q=(16/15)(2/\pi)^{1/2}[\Gamma(7/4)]^2\approx 0.719. \eeqno{kuzmin}
\par
A family of finite galactic disks has been described in Ref. \cite{brada94} having a Newtonian RC $V_N(r)= \eta u^\a$ \{$u\equiv r/r\_0$, $\eta=[\pi^{1/2}(2\a+1)\Gamma(\a+1/2)/2\Gamma(\a+1)]^{1/2}
(MG/r\_0)^{1/2}$\} within the material disk, whose surface density vanishes at and beyond $r\_0$, and inside $r\_0$: \beq\S_{\a}(r)=\frac{(2\a+1)M}{2\pi r^2\_0} v(1-v^2)^{(\a-1)}{}_2F_1(1-\a,\frac{1}{2};\frac{3}{2};v^2), \eeqno{mioop}
where $v\equiv (1-u^2)^{1/2}$, and $_2F_1$ are hypergeometric functions. For these, I find
\beq \Q=\left[\frac{\pi^{3/2}(2\a+1)}{18}
\frac{\Gamma(\a+1/2)}{\Gamma(\a+1)}\right]^{1/2}
 \frac{\Gamma(\a/2+5/4)}{\Gamma(\a/2+7/4)}.  \eeqno{maipo}
As $\a$ varies in the relevant range\footnote{Disks with $\a<1/2$, such as the Mestel disk (with $\a=0$, for which $\Q\approx 0.730$) are not legitimate DML disks, because their accelerations diverge at the center. For $\a>3/2$ the MOND RC is (unrealistically) concave.} $1/2\le\a\le 3/2$, $\Q$ varies only between 0.740 and 0.726.
Some interesting special cases are the `isodynamic' disk \cite{milgrom89c}, which has a constant acceleration (both in Newtonian and MI DML), with $\a=1/2$, for which $\Q=\pi/2^{1/2}3\approx 0.740$; the Kalnajs disk, with $\a=1$, for which $\Q\approx 0.736$; and $\a=3/2$, corresponding to rigid rotation in MOND, for which $\Q=2^{7/2}/3^{5/2}\approx 0.726$.
\par
For all exponential disks, which have $\S(r)\propto e^{-r/h}$, one has
 \beq \Q=8\int_0^\infty s^{5/2}e^{-2s}[I_0(s)K_0(s)-I_1(s)K_1(s)]^{1/2}ds,\eeqno{tarata}
where $I_\n$ and $K_\n$ are the modified Bessel functions of the first and second kind ($s=r/2h$). Evaluated numerically, $\Q\approx 0.733$.
\par
To expand the range of disk models, I also considered double exponentials with
 \beq \S(r)\propto e^{-r/h}+\b e^{-qr/h},  \eeqno{exaexa}
 with the two disks having a mass ratio of $\b/q^2$.
 For these, $\Q$ can be written as
$$\Q=\frac{8q^2}{q^2+\b}\int_0^{\infty}s^{5/2}(e^{-2s}+\b e^{-2qs}) \{I_0(s)K_0(s)$$
\beq -I_1(s)K_1(s)+\b q[I_0(qs)K_0(qs)-I_1(qs)K_1(qs)]\}^{1/2}ds.  \eeqno{mlok}
For the pairs $(q,\b)=(2,1),~(5,1),~(2,4),~(3,9)$ (the last two correspond to equal masses), I find $\Q\approx 0.7312,~0.7319,~0.7299,~0.7257$.
Taking $-1\le\b<0$, we get exponential disks suppressed near the center (an actual hole occurs if $q|\b|>1$, but if $q|\b|$ is too large $V_N^2$ becomes negative).
For the pairs $(q,\b)=(2,-0.5),~(1.5,-0.7),~(1.2,-0.8),~(2,-0.55),~(4,-0.3)$, $\Q\approx
0.7340,~0.7346,~0.7346,~0.7341,~0.7332$.
\par
Surprisingly, and for reasons that I do not understand, all the above disk models (which are all that I tried) give very near values of $\Q=0.73\pm 0.01$, compared with the universal MG value of $\Q=2/3$. The difference is small, amounting to about a 5\% difference in the rms velocity, indicating that the predicted RCs in all the theories concerned differ only a little from each other.\footnote{Looking at the few comparisons of full RCs in Ref. \cite{brada95} we see that the contribution to the difference in $\Q$ comes from the bulk of the disk (roughly around where $\S$ takes half its central value).} Using $\Q$ as discriminator may be rather demanding, considering the possible sources of systematic errors (see section \ref{discussion}). Still, hopefully, with accurate enough data for a large enough sample of galaxies, these differences might be used to distinguish between the two classes of theories.

\section{\label{discussion}Discussion}
Since the differences in rms velocities predicted by the two theory classes are only about 5\%, the proposed test will be potentially stymied by measurement and various systematic errors. The main concerns I can think of are as follows:
(i) Accurate determination of $\vinf$ for low-acceleration galaxies is not always possible, since their speed at the furthest observed radius is, typically, larger than $\vinf$ by a few percent. For example, the predicted DML RC of an exponential disk peaks at about 5 scale lengths, beyond which it declines to $\vinf$, which is about 5\% lower. This fact has been discussed in Ref. \cite{mcgaugh11a} in connection with its effect on the measured intercept of the $M-\vinf$ relation [Eq.(\ref{plama})].
(ii) There may be a possible unaccounted for contribution to the integral in $\Q$ from radii beyond the furthest observed one. Since $V\approx \vinf$ there, the missing contribution to $\Q$ is the fractional disk mass beyond this radius.
(iii) In very low velocity galaxies, which some low-acceleration disks are, the analysis is susceptible to uncertain, asymmetric-drift corrections. It is thus advisable to exclude such galaxies from the analysis. (iv) In determining $\Q$ observationally, we need the observed RC, as well as $\S(r)$
(both only up to an immaterial normalization). If the disk is made of gas only, or of stars only (for which we may assume a constant $M/L$ ratio) this can be cleanly done. But if both components contribute, the exact shape of $\S(r)$ depends on the stellar $M/L$ value, which is not known a priori with enough accuracy. It is thus advisable to use galaxies that are dominated by either component (as used in Ref. \cite{mcgaugh11b} for a clean test of the MOND relation Eq. (\ref{plama})).
\par
Other possible systematics to worry about are, departures from the DML , and the applicability of the razor-thin-disk approximation, assumed all along.
\par
We may also consider other moments of the velocity curves as discriminants. While they may have some advantages, they will have the drawback that there is no simple expression for them (not even as integrals) in MG theories, and they will have to be computed for each surface density law separately (for MI theories they can be calculated with the same ease as $\Q$). For example, I looked at $\Q_4\equiv\langle V^4\rangle^{1/2}$.
For Kuzmin disks, $\Q_4$ can be calculated analytically for both theory classes. I find $\Q_4=(8/15)^{1/2}\approx 0.730$ for the two MG theories, and $\Q_4=(3\pi/16)^{1/2}\approx 0.767$ for all MI theories. These are a little larger [as expected since $V(r)$ is increasing], and nearer each other than the $\Q$ values [as expected since the predicted $V(r)$ of all theories merge asymptotically].

\clearpage


\begin{thebibliography}{}
\bibitem{fm12}B. Famaey and S. McGaugh,  Living Rev. Relativity, 15, 10, arXiv1112.3960 (2012).
\bibitem{bm84}J. Bekenstein and M. Milgrom, Astrophys. J. 286, 7 (1984).
\bibitem{milgrom10a}M. Milgrom, Mon. Not. R. Astron. Soc. 403, 886 (2010).
\bibitem{milgrom94a}M. Milgrom, Ann. Phys. 229, 384 (1994).
\bibitem{milgrom11}M. Milgrom, Acta Physica Polonica B vol. 42, 2175 (2011).
\bibitem{bek04}J.D. Bekenstein, Phys. Rev. D 70, 083509 (2004).
\bibitem{zlosnik07}T.G. Zlosnik, P.G. Ferreira, and G.D. Starkman, Phys. Rev. D 75 044017 (2007).
\bibitem{milgrom09}M. Milgrom, Phys. Rev. D 80, 123536  (2009).
\bibitem{blt09}L. Blanchet, and A.  Le Tiec,
Phys. Rev. D 80, 023524 (2009).
\newpage
\bibitem{milgrom09a}M. Milgrom, Astrophys. J. 698, 1630 (2009).
\bibitem{brada95}R. Brada and M. Milgrom, Mon. Not. R. Astron. Soc. 276, 453 (1995).
\bibitem{angus12}G.W. Angus, K. van der Heyden, B. Famaey, G. Gentile, S.S. McGaugh, and W.J.G. de Blok, Mon. Not. R. Astron. Soc., 421, 2598 (2012)
\bibitem{milgrom94b}M. Milgrom, Astrophys. J. 429, 540 (1994).
\bibitem{brada94}R. Brada and M. Milgrom, Astrophys. J. 444, 71 (1994)
\bibitem{milgrom89c}M. Milgrom, Astrophys. J. 338, 121 (1989).
\bibitem{mcgaugh11a}S. S. McGaugh, Phys. Rev. Lett. 107, 229901(E) (2011).
\bibitem{mcgaugh11b}S. S. McGaugh, Phys. Rev. Lett. 106, 121303 (2011).

\end{thebibliography}
\end{document}